# Leveraging Recurrent Patterns in Graph Accelerators

Masoud Rahimi, Sébastien Le Beux
Department of Electrical and Computer Engineering, Concordia University, Montreal, Canada
Email: masoud.rahimi@mail.concordia.ca, sebastien.lebeux@concordia.ca

*Abstract*— Graph accelerators have emerged as a promising solution for processing large-scale sparse graphs, leveraging the in-situ computation of ReRAM-based crossbars to maximize computational efficiency. However, existing designs suffer from memristor access overhead due to the large number of graph partitions. This leads to increased execution time, higher energy consumption, and reduced circuit lifetime. This paper proposes a graph processing method that minimizes memristor write operations by identifying frequent subgraph patterns and assigning them to graph engines, referred to as *static*, allowing most subgraphs to be processed without a need for crossbar reconfiguration. Experimental results show speed up to 2.38× speedup and 7.23× energy savings compared to state-of-the-art accelerators. Furthermore, our method extends the circuit lifetime by 2× compared to state-of-the-art ReRAM graph accelerators.

## I. INTRODUCTION

Graphs are a fundamental tool for modeling and analyzing relationships across various application domains, such as social networks [1], recommendation systems [2], [3], computer networks, and bioinformatics [4]. For instance, in social networks, users are modeled as nodes and their connections as edges to analyze interactions and influence. Real-world graphs often contain millions of nodes [5], [6], [7], are typically sparse and characterized by unstructured connectivity. Processing such graphs introduces several challenges, including: i) large memory capacity requirement, ii) random memory accesses caused by irregular connections, and iii) complexity of processing irregular graph structures.

To address these challenges, memory-centric architectures have been proposed to accelerate graph processing [8]. These architectures typically utilize in- or near-memory processing paradigms to reduce memory access [9]. Given the large size and inherent sparsity of graphs, conventional accelerators partition graphs into smaller subgraphs to satisfy limited capacity of local memories [10]. These subgraphs are then mapped onto ReRAM-based crossbars, where computation is performed directly within crossbars [10], [11], [12]. In these methods, subgraph data are stored in adjacency matrix format. To mitigate inefficiencies introduced by large-scale and sparsity of real-world graphs, several methods have been proposed to improve performance. For instance, ReGraphX [13] utilizes relatively small crossbars (e.g., 8×8) for edge-centric processing in Graph Neural Networks (GNNs), reducing storage of zero entries. Spara [14] introduces a graph reordering and vertex remapping scheme in preprocessing to maximize crossbar utilization. Alternatively, SparseMEM [15] proposes a hierarchical mapping scheme that mitigates local crossbar underutilization and enables core graph computations on compressed graph representations. ReFlip [12] introduces a flipped-mapping scheme designed to increase parallelism of Matrix-Vector-Multiplication (MVM) operations in crossbars. In this design, feature vectors of a graph are mapped onto the crossbars, while edge data are fed as inputs. Overall, existing methods mainly address graph sparsity to improve graph accelerator performance by reducing memory constraint requirements.

A pattern is defined by the adjacency matrix of a graph, where '1' represents an edge between two vertices, and '0' indicates absence of an edge. After partitioning a graph, the pattern occurrence refers to the number of subgraphs sharing an identical pattern. Figure 1-a reports pattern occurrence in Wiki-Vote [5], an unweighted graph with 7K vertices and 104K edges. The data are obtained by applying a 4×4 non-overlapping sliding window on the adjacency matrix, a commonly used partitioning approach [13], [16]. Pattern with all '0' is discarded since it does not involve any processing. As reported in the figure, the most frequent pattern $P_0$ accounts for 5.9% of subgraphs and, altogether, the 16 mostly frequent ($P_0$ to $P_{15}$) account for 86% of subgraphs. In contrast, remaining patterns ($P_{16}$ to $P_{809}$) collectively account for only 14% of subgraphs. Similar distributions were observed across all studied datasets [5], [6]. By leveraging the observation that few patterns allow processing most subgraphs, graph accelerators can be designed with minimal ReRAM write operations, that are known to be energy consuming, slow, and impact memory cells lifetime.

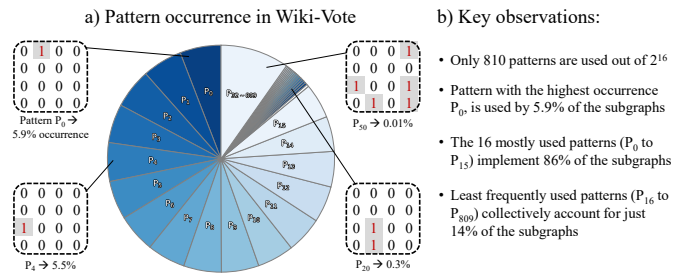

Figure 1: a) Pattern occurrence in Wiki-vote [5] partitioned with a 4×4 sliding window on the graph's adjacency matrix. b) Key observations.

In this paper, for the first time, we leverage pattern occurrence to design graph accelerators. To this end, we develop a method that identifies and ranks patterns based on their frequency in the input graph. The most frequent patterns are then statically mapped to graph engines, which leads to a drastic reduction of memory crossbar writes. The proposed method is fully automated and consists of: i) a window-based partitioning algorithm to preprocess the input graph, ii) a scheduling algorithm capable of executing classical graph algorithms while handling graph engine configuration at run-time, and iii) a design space exploration framework that identifies optimal architectural parameters. Experimental results demonstrate that our architecture achieves up to 2.38× speedup and 7.23× energy savings compared to state-of-the-art accelerators, while also increasing circuit lifetime by up to 2x thanks to reduced ReRAM writes.

The rest of the paper is organized as: Section II presents background and related work. Section III presents the proposed method. Section IV presents our experimental setup and results. Finally, Section V concludes the paper.

## II. BACKGROUND AND RELATED WORK

### A. ReRAM-based Crossbars

Resistive Random Access Memory (ReRAM) is a non-volatile emerging memory based on a metal–oxide–metal structure where data are stored by altering the resistance states of individual cells [17]. ReRAM offers fast read speeds and high storage density compared to other non-volatile memories [18], [19]. In contrast, writes are slower than reads and consume more energy [20], [21]. Moreover, ReRAM devices have limited endurance, between $10^5$ to $10^8$ write cycles [22], [23]. ReRAM cells are arranged in crossbar arrays [24], enabling in-situ MVM. Each bitline performs a MAC operation expressed as $\sum_i G_{i,j}.V_i$, where $G_{i,j}$ denotes the conductance and $V_i$ is the input voltage applied to wordlines [10], [25]. Therefore, a crossbar efficiently performs MVM in $O(1)$ computational complexity.

### B. Graph Partitioning and Mapping to ReRAM

Realistic graph applications typically involve millions of vertices and edges [5], [6], [7]. Processing such large graphs is challenging due to memory capacity limitations. To address this, the graph is partitioned into smaller subgraphs using partitioning methods. For instance, the method detailed in [26] involves duplicating vertices across multiple subgraphs, and distributing edges among them. This allows precise control over the number of vertices within each subgraph. Vertex duplication is particularly effective for compressed graph representations, where only edges are stored. Window-based partitioning is another method for graphs represented in adjacency matrix format. In this approach, a non-overlapping sliding window is applied across the matrix to divide it into smaller submatrices and are directly mapped onto ReRAM-based crossbars [11]. Since submatrices containing only zeros are excluded, using a small window size can significantly reduce storage requirements for sparse graphs [10]. Our method relies on window-based partitioning, using small windows (e.g., 4x4) to maximize zero submatrices, reducing memory footprint.

To store graph data in crossbars, both compressed and uncompressed formats have been explored. For instance, Spara [14] and Janus [27] utilize Compressed Sparse Row (CSR) to improve memory utilization. This representation enables efficient access to edges of active vertices, allowing simultaneous traversal from a common source. In contrast, GraphR [10] and Remagn [11] store graphs using uncompressed adjacency matrix, enabling in-situ MVM on ReRAM crossbars. In our design, we store input graphs using Coordinate list (COO) format. This ensures efficient storage and sequential edge access, while utilizing adjacency matrix format in local memory to enable in-memory processing on ReRAM.

### C. Graph Accelerators

Prior studies have leveraged in-situ MVM, a fundamental operation in many graph algorithms to accelerate graph processing [10], [12], [13]. For example, GraphR [10] achieves up to two orders of magnitude speedup over a CPU baseline for classical graph algorithms. ReGraphX [13] proposes heterogenous crossbars for on-chip GNN training. However, large-size and sparsity of real-world graphs combined with limited number of graph engines necessitate frequent memory access (i.e., read and write operations) in ReRAM-based crossbars. Consequently, this leads to i) increased execution time, ii) high energy consumption, and iii) reduced circuit lifetime due to limited ReRAM endurance.

Recent works mitigate inefficiencies caused by graph sparsity, as summarized in Table 1. For instance, GraphIte [28] applies graph sparsification by selectively removing edges with minimal impact on vertex updates at the expense of some accuracy loss. Alternatively, SparseMEM [15] proposes a compressed hierarchical mapping analogous to CSR format. In this design, destination vertices and associated weights are stored sequentially within one crossbar, while vertex locations are stored in a separate crossbar. This maximizes crossbar utilization by eliminating zero-valued cells but precludes in-situ MVM operations and requires high-resolution Multi-Level Cell (MLC) ReRAM to store vertex indices. TARe [16] introduces a write-free mapping strategy that partitions a ReRAM crossbar into multiple smaller computing blocks (CBs), each preconfigured with complete sets of possible binary submatrices. While this eliminates runtime write operations, it restricts parallel MVM operations and incurs frequent off-chip memory reads, degrading performance. In contrast, our design minimizes ReRAM writes without excessive off-chip memory access and operates on 1-bit ReRAM crossbars.

Table 1: Comparison of existing graph accelerators

| Reference | In-engine Graph Representation | Memory Access (Read/Write) | Required MLC[1] ReRAM | Graph Algorithm |
|---|---|---|---|---|
| GraphR [10] | Adjacency | High/High | 4-bit | Classical |
| ReFlip [12] | Compressed | High/Low | Variable[2] | GNN |
| SparseMEM [15] | Compressed | Low/Low | Variable[3] | Classical |
| TARe [16] | Adjacency | High/Low | 1-bit | GNN |
| Proposed | Adjacency | Low/Low | 1-bit | Classical |

1: Multi-Level Cell. 2: Depends on the size of input feature vectors. 3: Depends on the number of vertices.

## III. PROPOSED METHOD

In this section, we first present the design flow used to configure the proposed architecture. We then introduce the preprocessing model, which analyzes the input graph and extracts structural patterns. Next, we present the graph processing algorithm. Finally, we detail the graph engine architecture.

### A. Design Flow

Figure 2 presents our design flow, which takes as inputs a graph and an architecture model. The architecture is generic and is characterized by the following parameters: crossbar size (C), total number of graph engines (T), number of static graph engines (N), and number of crossbars per graph engine (M).

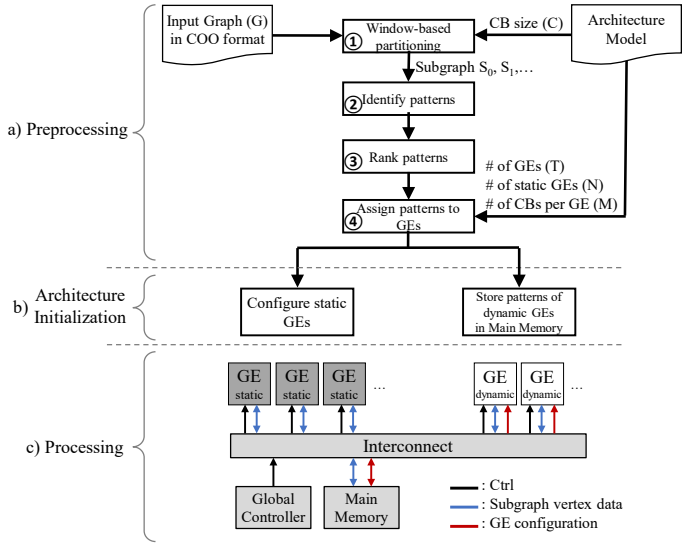

Figure 2: Design flow and architecture overview: a) input graph preprocessed based on architectural parameters; b) architecture initialized; and c) preconfigured architecture with static and dynamic graph engines.

During preprocessing (shown in Figure 2 upper part); ① the graph is partitioned into subgraphs sized by crossbar size. Patterns are then identified per subgraph② and ranked by frequency across the graph ③. In ④, the most frequent patterns are assigned to static graph engines, which are configured once. The least frequent patterns are assigned to dynamic graph engines, reconfigured at runtime. Section III.B details our preprocessing algorithm. The architecture (depicted at the bottom of Figure 2) is composed of multiple graph engines to enable parallel processing. To maximize graph engine activations, most engines are implemented as static and configured once during the initialization phase. In contrast, dynamic engines are reconfigured as needed at runtime.

### B. Graph Preprocessing Algorithm

Algorithm 1 outlines preprocessing where the inputs are graph $G$ and architectural parameters. Outputs are configuration table $CT$, and subgraph table $ST$. The algorithm: i) partition input graphs into smaller subgraphs based on the crossbar size $C$ (Line 4), ii) extract patterns from subgraphs and rank them by frequency (Lines 5–12). iii) assign most frequent patterns to static engines and map remaining patterns to dynamic engines (Lines 13–19).

Figure 3 illustrates preprocessing of a graph with six vertices using three graph engines, each with a single 2×2 crossbar. The graph is converted to an adjacency matrix, partitioned into nine subgraphs ($S_0$ to $S_8$), then patterns are extracted (Figure 3-b) and ranked by frequency (Figure 3-c). For example, pattern $P_0$ has the highest frequency, appearing in $S_0$, $S_4$, and $S_6$. In contrast, $P_2$ and $P_3$ each appear once, in $S_3$ and $S_5$, respectively. As mentioned earlier, subgraphs with no edges ($S_5$ and $S_8$), corresponding to patterns consisting entirely of zeros, are excluded from the ranking.

Following pattern identification and ranking, patterns are mapped to graph engines based on their frequency (Figure 3-d). Most frequent patterns, $P_0$ and $P_1$, are assigned to static engines $GE_0$ and $GE_1$, respectively, while less frequent patterns, $P_2$ and $P_3$, are allocated to the dynamic engine $GE_2$. Finally, patterns and subgraphs are stored in the main memory (Figure 3-e). The configuration table stores the pattern data, represented in COO format, along with their associated graph engines. The subgraph table stores vertex data for each subgraph along with its corresponding pattern. Since all subgraphs have same number of vertices, only the starting source and destination vertices are recorded. For instance, for $S_1$, only vertices $V_0$ and $V_2$ are stored, thereby reducing storage overhead in main memory.

---

**Algorithm 1:** Preprocessing

**Input:** Graph $G$, crossbar size $C$, No. of static graph engines $N$, No. of crossbars per graph engine $M$

**Output:** Graph engine configuration table $CT$, subgraph table $ST$

1 Let **Partitioning(Graph, C):** partition *Graph* into subgraphs with $C$ vertices.
2 **Procedure** Preprocess($G$, $C$, $N$, $M$):
3     $patterns \leftarrow \{\}$
    // partition the input graph
4     $subgraphs \leftarrow$ Partitioning($G$, $C$)
    // identify & rank patterns
5     **foreach** $s \in subgraphs$ **do**
6        $p \leftarrow$ GetPattern($s$)
       // increment frequency for an existing pattern
7        **if** $p \in patterns$ **then**
8           $patterns[p] \leftarrow patterns[p] + 1$
9        **else**
10           $patterns$.add($p$, 1)
11     $ST$.add($s$, $p$)
    // sort patterns by frequency
12     $patterns$.sort()
    // assign patterns to graph engines
13     $i \leftarrow 0$
14     **foreach** $p \in patterns$ **do**
15        **if** $i < N * M$ **then**
16           $CT$.add($p$, "static")
17           $i \leftarrow i + 1$
18        **else**
19           $CT$.add($p$, "dynamic")
20     **return** $CT$, $ST$

---

In case graph engines contain multiple crossbars (i.e., $M > 1$), patterns assigned to static engines are evenly distributed across their crossbars (function $FindGE$ in algorithm 1). This distribution balances pattern load among static engines, improving overall utilization. Furthermore, since patterns with a single edge are more frequent (due to power-law degree distribution [29]), the row address is also stored in the configuration table. This eliminates iteration over all crossbar rows, thereby reducing ReRAM reads in static engines.

### C. Graph Processing

For runtime processing, we use the streaming-apply execution model detailed in [10]. In this model, subgraphs sharing same source vertices (row-major order) or destination vertices (column-major order) are processed together within an iteration. The table in Figure 3-e shows subgraphs in column-major order. For example, $S_0$, $S_3$, and $S_6$ are grouped together since they share same destination vertices $V_0$ and $V_1$. Our design supports both row- and column- major execution models, which can be selected for different graph algorithms. Furthermore, execution order of subgraphs can be further optimized to maximize engine activations, which we leave as future work.

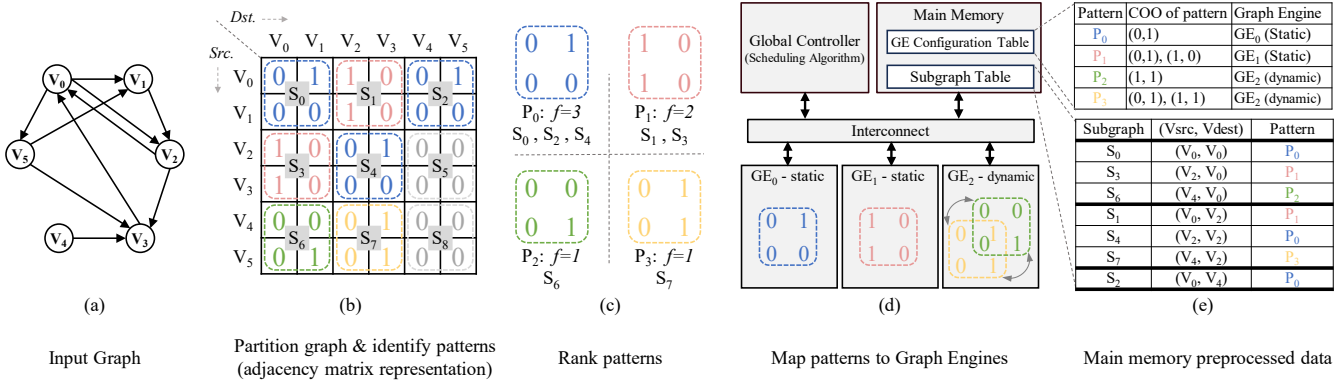

Figure 3: Graph preprocessing: a) Input graph, b) partition and identify patterns in adjacency matrix representation, c) rank patterns, d) map to two static and one dynamic graph engines, and e) graph engine configuration and subgraph tables (column-major order) in main memory.

Algorithm 2 details our graph processing. Static graph engines are pre-configured with assigned patterns. Subsequently, subgraphs are processed over multiple iterations, where in each iteration, a batch (a group of subgraphs) is fetched from the subgraph table (*ST*) (Line 9) and executed in parallel (Line 10). Batch size is determined by available graph engines. For each subgraph in a batch, if the pattern is assigned to a static engine, only vertex data is transferred (Line 12). Otherwise, a dynamic engine is selected based on the replacement policy (*FindGE* function) and then reconfigured with the corresponding pattern (Line 14-15). Once all subgraphs in a batch are processed (*Allocate* function), the updated vertex data is aggregated (Line 17).

```
Algorithm 2: Graph Processing & Scheduling
  Input: Graph engine configuration table CT, subgraph table ST
  Output: Processed Vertex Data Results
1 Let Configure(GE, Config) configure graph engine GE with Config.
2 Let Allocate(GE, VertexData) allocate VertexData to graph engine GE.
3 Let Aggregate(VertexData) aggregate VertexData from the graph engines.
4 Procedure Process(CT, ST):
5     Results ← {}
      // initialization: configure static GEs
6     foreach pattern p ∈ CT do
7         if p is static then
8             Configure(p.GE, p.data)
      // iteratively process subgraphs
9     foreach batch of subgraphs with same dest. vertices ∈ ST do
10        parallelforeach subgraph s, pattern p ∈ batch do
11            if p is static then
                  // process subgraph with initial vertex data
12                r ← Allocate(p.GE, s.vertex_data)
              // process subgraph in dynamic GE
13            else
14                ge ← findGE(p.data)
15                Configure(ge, p.data)
16                r ← Allocate(ge, s.vertex_data)
              // aggregate vertex data from GEs
17            Results ← Aggregate(r)
18    return Results
```

### D. Graph Engine

Figure 4 illustrates the architecture, with two ReRAM crossbars: $CB_0$ (static) and $CB_1$ (dynamic). Graph execution is managed by the control unit, which handles crossbar configuration, data movement, and scheduling. The input register holds initial vertex data, while the output register stores the processed results. Analog outputs from crossbar bitlines are sampled by sample-and-hold (S/H) circuits and digitized by analog-to-digital converters (ADCs), which are shared across multiple bitlines to reduce area and power overhead. The driver applies input voltages during MVM operations and configures crossbars as needed. Computations not natively supported by in-situ MVM are offloaded to a lightweight arithmetic logic unit (ALU). Note that the architecture shown is generic, however our method requires only static or dynamic crossbars within an engine. Furthermore, we assume same crossbar size for both static and dynamic engines. We leave exploration of non-uniform crossbar configurations to future work.

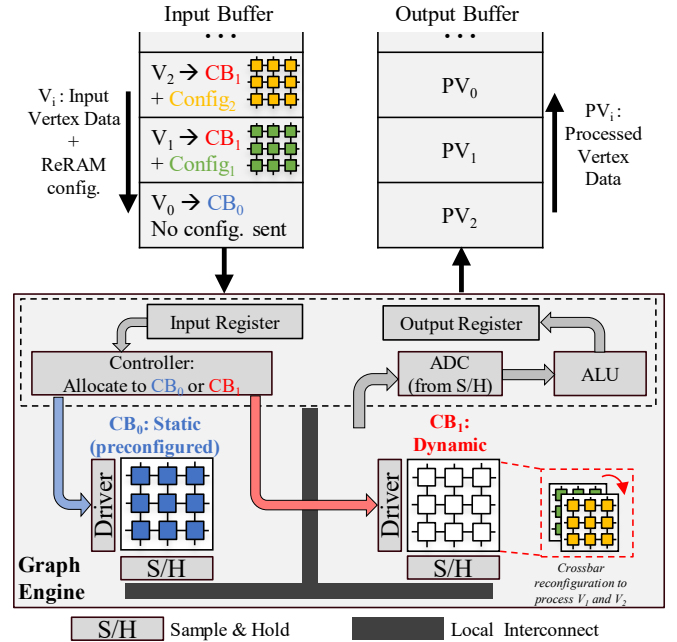

Figure 4: Graph engine architecture composed of two crossbars.

At runtime, the vertex data ($V_i$) of a given subgraph is loaded from the input buffer. For dynamic crossbars (e.g., $CB_1$ in Figure 4), the crossbar configuration (edge data labeled as $Config_i$) is also transferred via the input buffer. In contrast, for static crossbars (e.g., $CB_0$), no configuration data is required. After processing, updated vertex data ($PV_i$) is written back through an output buffer. Both input and output buffers operate as FIFOs, where each input entry is paired with a corresponding output entry, enabling pipelined processing of multiple subgraphs.

Our architecture supports a range of graph algorithms such as Breadth-First Search (BFS), Single-Source Shortest Path (SSSP), and PageRank that follow the vertex programming model described in [10]. In this model, graph processing is divided into two phases: i) *edge computation*, where a value is computed for each outgoing edge of an active vertex using in-situ MVM operation of ReRAM crossbars; and ii) *reduce and apply*, where each vertex property (i.e., vertex data) is updated by applying a reduction function over all incoming edge values using the ALU.

## IV. EVALUATION

In this section, we first present the experimental setup used to evaluate our design. We then analyze activity of graph engines to demonstrate the functionality of our method and identify optimal architectural parameters. Finally, we compare our design performance against state-of-the-art accelerators.

### A. Experimental Setup

We develop a system-level simulator to evaluate the design performance. It estimates the execution time and energy consumption by monitoring memory access performed by the graph engines during processing. To determine required number of clock cycles, the simulator models the parallel operation of graph engines. Several preprocessing parameters are provided to the simulator, including crossbar size, total number of graph engines, number of static and dynamic engines, and number of crossbars per engine. For comparison with state-of-the-art, we use the same crossbar configuration and peripheral circuitry.

We evaluate our design using datasets sourced from various domains, as listed in Table 2 [5]. These benchmarks are undirected, weighted graphs with varying sizes and sparsity levels, enabling a comprehensive analysis. The average degree of a graph represents the average number of edges per vertex, indicating the graph's connectivity level. For benchmarking, since input graphs are unweighted, we use BFS as the baseline graph algorithm. We also use column-major order as our baseline execution model [10].

Table 2: Graph datasets

| Name | #vertices | #edges | Average Deg. | Sparsity (%) | Domain |
|---|---|---|---|---|---|
| web-Google (WG) | 875K | 5.1M | 12 | 99.999 | Web |
| Amazon302 (AZ) | 262K | 1.2M | 9 | 99.998 | Recom. |
| Slashdot0902 (SD) | 82K | 948K | 23 | 99.985 | Social |
| soc-Epinions1 (EP) | 76K | 509K | 13 | 99.991 | Social |
| p2p-gnutella31 (PG) | 5K | 148K | 5 | 99.996 | Network |
| Wiki-vote (WV) | 7K | 104K | 29 | 99.795 | Social |

Unless stated otherwise, we assume 32 graph engines containing 4×4 crossbars. We consider single-level cell (i.e., 1-bit) ReRAM crossbars, modeled in 32nm node technology, with capacity of 32KB and 8 bits data width. ReRAM parameters are obtained from NVSim [30]. We use CACTI-6.5 [31] at 32nm to simulate I/O buffers and main memory. Furthermore, ADC parameters are derived from [32], assuming a 32 nm technology node with 8-bit resolution. Table 3 summarizes the simulation specifications.

Table 3: Simulation specifications

| | | Latency | Energy |
|---|---|---|---|
| 4x4 ReRAM crossbar 32KB $V_{SET} = V_{RESET} = 2V$ | Per bit Read | 1.3ns | 1.1pJ |
| | Per bit Write | 20.2ns | 4.9pJ |
| | Sense Amplifier | 1ns | 1pJ |
| SRAM buffer 32KB | Per access | 0.31ns | 29pJ |
| ADC 8-bit resolution | Per access | 1ns | 2pJ |

In our evaluation, we consider both graph engine performance and main memory access. Since baseline designs perform better with large crossbars, we assume a 128×128 crossbar configuration with the same capacity. For TARe [16], which is primarily designed for GNN applications, we consider only its mapping scheme and adapt it for classical graph algorithms.

### B. Graph Engine Activity

Analyzing graph engine activity characterizes resource utilization, workload distribution, and potential performance bottlenecks. For this purpose, we assume a design with 6 graph engines including 4 static and 2 dynamic, each containing 4 crossbars. Figure 5 illustrates the crossbar read and write activity during processing of Wiki-Vote. Activity levels (scale of 0 to 100) are computed by aggregating number of crossbar reads/writes over a sliding window of iterations. We observe non-uniform read and write activity across iterations. This is caused by the execution model, where the number of subgraphs processed in parallel varies across iterations. Analysis of other datasets shows this behavior is strongly graph-dependent. Since most subgraphs are processed by static engines, their read activity is significantly higher than dynamic engines. This variation motivates us to explore workload balance between static and dynamic engines to achieve higher speedup.

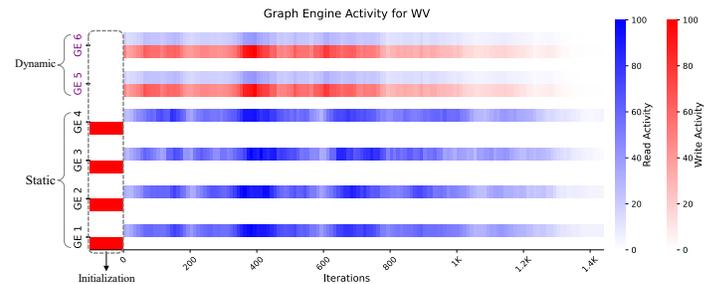

Figure 5: Graph Engine (GE) activity during processing of Wiki-Vote. $GE_1$ to $GE_4$ are static, while $GE_5$ and $GE_6$ are dynamic. Activity levels range from 0 (no activity) to 100 (maximum activity).

We now study the trade-off between number of static and dynamic graph engines, keeping the total number fixed. Figure 6 shows speedup achieved for different allocations of static engines, normalized to the baseline configuration with no static engines. For simplicity we assume a single crossbar per engine. For the sake of brevity, results are shown for three representative datasets. Across all three datasets, assigning 16 static engines results in highest speedup. For instance, in WS, utilizing 16 static engines achieves a 1.8× speedup. Given a 4×4 crossbar, most subgraphs contain only a single edge due to the power-law degree distribution [29]. Therefore, at least 16 static engines are required to ensure all such subgraphs are processed by static engines. In contrast, assigning most engines as static leads to longer execution time because few dynamic engines limit parallelism and increase the iteration count. We observed a similar trend across other datasets. In the following, we use the activity profile to quantify total memory access and evaluate system performance.

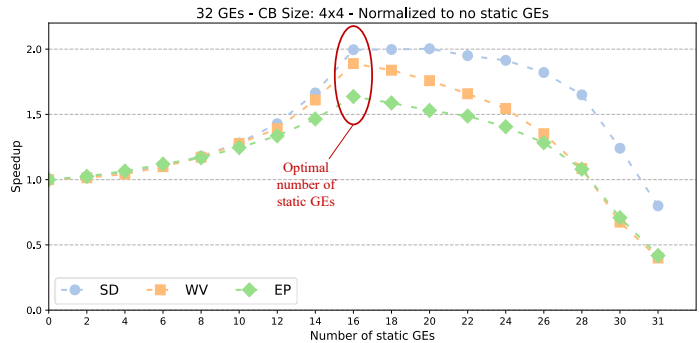

Figure 6: Speedup wrt. to the number of static graph engines. A total of 32 graph engines, each containing a 4×4 crossbar. Results are normalized to a system with no static engines.

### C. Performance Comparison

In the following, we compare the performance and energy consumption of our design with three baseline approaches GraphR [10], SparseMEM [15], and TARe [16]. For fair comparison, we assume equal number of graph engines and identical memory capacity across all designs.

*1) Energy*: Table 4 reports the energy consumption of our proposed design and state-of-the-art methods for executing the BFS algorithm on each dataset. Overall, our design is three orders of magnitude more energy efficient than GraphR and achieves 7.23× and 2.3× energy efficiency compared to SparseMEM and TARe, respectively. The improvement over GraphR and SparseMEM is explained by fewer ReRAM writes. For TARe, although the write-free engine suppresses ReRAM writes, frequent off-chip memory accesses lead to higher overall energy consumption.

Table 4: Total energy consumption of BFS across evaluated datasets.

| Dataset | GraphR [10] | SparseMEM [15] | TARe [16] | Proposed |
|---------|-------------|----------------|-----------|----------|
| WG | 4.1 J | 2.12 mJ | 470 μJ | 318 μJ |
| AZ | 460 mJ | 688 μJ | 79 μJ | 54 μJ |
| SD | 110 mJ | 260 μJ | 50 μJ | 48 μJ |
| EP | 53 mJ | 182 μJ | 35 μJ | 26 μJ |
| PG | 60 mJ | 55 μJ | 30 μJ | 7.1 μJ |
| WV | 3.3 mJ | 23 μJ | 24 μJ | 5.9 μJ |

*2) Speedup:* Figure 7 compares speedup of the four designs. On average, our design achieves three orders of magnitude higher speedup than GraphR, and 2.38× and 1.27× over SparseMEM and TARe, respectively. The improvement over GraphR results from significantly reduced crossbar access, as GraphR performance is constrained by sparse subgraph mapping. Although SparseMEM achieves better crossbar utilization, its execution time is higher due to decompression of graph data in graph engines. Finally, while TARe minimizes ReRAM writes by configuring crossbars only once, it limits parallel execution of MVM operations, resulting in more iterations.

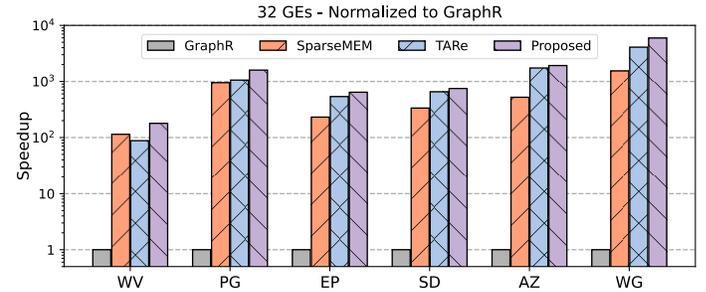

Figure 7: Speedup of the proposed design compared to the baseline. Results are obtained for BFS algorithm and normalized to GraphR.

### D. Circuit Lifetime Analysis

ReRAM cells have limited write endurance (∼108 cycles [23]), and exceeding this limit can cause write-induced faults [17], [33]. We model circuit lifetime as $\frac{E}{w} \times T$, where $E$ is cell endurance, $w$ is the maximum number of write operations per cell, and $T$ is the execution interval. We assume graph engines are not used once a crossbar reaches maximum writes, allowing remaining engines to continue operation. Furthermore, static engines are excluded, as they are configured only once. Considering 128 graph engines and executing Wiki-Vote once per hour ($T = 1h$), our design can operate for over 10 years, which is two orders of magnitude longer than GraphR and 2× longer than SparseMEM. The improvements are due to reduced number of ReRAM writes in our design.

## V. CONCLUSION

In this paper, we presented a graph accelerator that leverages recurring graph patterns to mitigate frequent memory access. We also propose a method to find the best number of static graph engines for a given application and to explore design tradeoffs. Our results show the proposed accelerator achieves 1.27× speedup and 2.3× reduction in energy consumption compared to state-of-the-art. As opposed to state-of-the-art graph accelerators that require expensive and error-prone, large ReRAM crossbars, our architecture performs better with smaller, cost-effective crossbars, e.g., 4x4 or 8x8. Our architecture thus has potential for deployment not only in HPC data centers, where graph accelerators are most needed, but also in high-performance embedded systems. Furthermore, static graph engines intrinsically do not require any ReRAM write operations during application execution, which significantly increases architecture lifetime. By leveraging graph remapping on graph engines, we plan to further enhance architecture reliability, which we will investigate in our future work.